\documentclass [12pt] {article}
\def\one{1\hskip-.37em 1}

\def\half{\textstyle{\frac{1}{2}}}
\def\quarter{\textstyle{\frac{1}{4}}}

\def\b{\begin{eqnarray*}}     
\def\e{\end{eqnarray*}}       
\def\bn{\begin{eqnarray}}     
\def\en{\end{eqnarray}}       
\def\<{\langle}
\def\>{\rangle}

\def\{{\lbrace}
\def\}{\rbrace}

\title{Generalized Affine Coherent States:\\
A Natural Framework for Quantization of Metric-like Variables}
\author{Glenn Watson and John R. Klauder\\
Departments of Physics and Mathematics\\
University of Florida\\
Gainesville FL 32611}
\date{}
\begin{document}
\maketitle

\begin{abstract}
Affine variables, which have the virtue of preserving the positive-definite character of matrix-like objects, have been suggested as replacements for the canonical variables of standard quantization schemes, especially in the context of quantum gravity.  We develop the kinematics of such variables, discussing suitable coherent states, their associated resolution of unity, polarizations, and finally the realization of the coherent-state overlap function in terms of suitable path-integral formulations.
\end{abstract}

\section{Introduction}\

It has been suggested that the \it affine \rm  group may have a significant role to play in a quantum theory of gravity.  The (one-dimensional) affine Lie algebra is in some sense the simplest possible non-abelian Lie algebra; it may be considered to be generated by two self adjoint operators $\kappa $ and $\sigma $ whose commutation relation reads
\bn
[\sigma , \kappa ] = i\sigma.
\en
There exists a representation of (\theequation) in which the spectrum of $\sigma$ is \it strictly positive\rm.  It has been noted by Klauder \cite{klauder1} that a multidimensional generalization of the algebra (\theequation) exists in which $\sigma$ is replaced by a matrix operator corresponding to a \it symmetric, positive definite matrix degree of freedom.  \rm Such an object is clearly well suited to the description of the spacial part of a metric tensor, and attempts have been made to construct an affine field theory in the context of quantum gravity.  It has been argued by Isham and Kakas \cite{isham1, isham2} that an affine algebra arises naturally in an attempt to quantize a nonlinear phase space such as that which exists in general relativity.  The strong-coupling limit of such a theory of gravity has previously been discussed by Pilati \cite{pilati1, pilati2, pilati3}.\

The coherent state representation associated with the one-dimensional affine algebra (\theequation) is fairly well known \cite{klauder-aslaksen, dkp}.  Our present aim is to generalize this analysis to a multi-dimensional affine algebra.  Our discussion is limited largely to kinematics; furthermore, we make no attempt to build a field theory.  These questions are considered by Klauder \cite{klauder2}.

\section{The One-Dimensional Affine Algebra and its Associated Coherent States}\

The algebra in (\theequation) generates a 2-parameter Lie group known as the \it affine group \rm (also known as the \it  ``$ax + b$" group \rm, with $a>0$) \rm - the group of \it translations and dilations of the real line\rm.  The operators $\kappa$ and $\sigma$ induce the following transformations:
\bn
e^{iB \kappa} \; \sigma \; e^{-iB \kappa} &=& e^{B} \; \sigma,\\
e^{-iF \sigma} \; \kappa \; e^{iF \sigma} &=& \kappa + F\sigma.
\en
Here $B$ and $F$ are real parameters.

It is well known that there exist three, faithful, inequivalent, irreducible unitary representations of the affine algebra, characterized respectively by the operator $\sigma$ posessing positive, negative, and null spectra \cite{GN, klauderskagerstam}.  The positive representation is of particular interest in this article since the positivity of $\sigma$ will be generalized to a positive definite \it matrix \rm degree of freedom (see section 5).  In this representation, the operators $\sigma$ and $\kappa$ are represented by the following operators:
\bn
\sigma &=& k,\\
\kappa &=& \; \half [\theta k + k\theta] \; ,
\en
with
\bn
\theta \; = \; -i \; \frac{\partial}{\partial k} \; ,
\en
where the representation space is the space of $L^2$ functions on the open interval $(0, \infty)$, equipped with the inner product
\bn
\<\phi | \psi \> &=& \int _0^\infty {\phi(k)^* \; \psi(k) \; dk},
\en
for any two functions $\phi$ and $\psi$ in the space.  The operator $\kappa$ acts to generate \it unitary dilations \rm in the representation space:
\bn
e^{-iB \kappa} \; \psi (k) &=& e^{-B/2}  \; \psi (e^{-B} k),\\
|\!| e^{-iB\kappa} |\psi \> |\!| &=& |\!| |\psi \> |\!|.
\en
It is important to note that $\theta$, unlike its canonical conjugate $\sigma$, is \it not \rm self adjoint in this representation - nor does it possess self-adjoint extensions.  Its interpretation as an observable is therefore not possible. 

We now define a family of unitary operators $U(F,B)$ via
\bn
U(F,B) = e^{iF\sigma}e^{-iB\kappa}.
\en
The composition rule for the operators in (\theequation) is
\bn
U(F', B') \; U(F,B) = U(F' + e^{-B'}F \; , \;  B'+B).
\en
This family of unitary operators may be used to constuct a set of coherent states,
\bn
|F,B\> = U(F,B)|\eta\>.
\en
Here $|\eta \>$ is an as yet unspecified normalized fiducial vector in the representation space.  The coherent states in (\theequation) admit a resolution of unity in the form
\bn
N^{-1} \; \int_{-\infty}^{\infty} dF \; \int_{-\infty}^{\infty} dB \; e^B \; |F, B\> \<F, B| &=& \one,
\en
the validity of which hinges on the fiducial vector admissibility criterion
\bn
N &=& 2\pi \int_0^{\infty} dk \; k^{-1} \; |\eta(k)|^2 \; < \; \infty.
\en
For a more general discussion regarding coherent states, see \cite{klauderskagerstam}.\

An example of a normalized admissible fiducial vector is provided by 
\bn
\eta_{1}(k) = C_{1}(\alpha, \beta) \; k^{\alpha} \; e^{-\beta k},
\en
where $\alpha$ and $\beta$ are positive real coefficients, and the normalization constant $C_{1}(\alpha, \beta)$ is given by 
\bn
C_{1}(\alpha, \beta) \equiv \frac {(2\beta)^{\alpha  \; + \; 1/2}}{\sqrt{\Gamma(2\alpha + 1)}}.
\en
It may be verified that this fiducial vector satisfies (14), with $N = 2\pi \beta / \alpha$.  The associated coherent state representative of a general vector $|\psi\>$ in the Hilbert space, $\<F, B|\psi \>$, exhibits the property that
\bn
\left[ie^{-B}\frac{\partial}{\partial F} \; -  \; \beta^{-1} \frac{\partial}{\partial B} \; - \; \beta^{-1}(\alpha + \half) \right] \<F, B|\psi \> = 0.
\en
The relation (\theequation) defines a complex {\it polarization} of the relevant function space, and its form is inherited directly from the property of the fiducial vector (15) described by
\bn
\left[ \sigma \; - \; i\beta^{-1}\kappa \; - \; \beta^{-1}(\alpha + \half) \right] |\eta\> = 0.
\en
This property, when re-written in the form
\bn
(\sigma - \<\sigma\>) \; |\eta\> &=& i\beta^{-1} \; (\kappa - \<\kappa\>) \; |\eta\>,
\en
where
\bn
\<\sigma\> &\equiv& \<\eta|\sigma|\eta\> \; \; = \; \; \beta^{-1}(\alpha + \half),\\
\<\kappa\> &\equiv& \<\eta|\kappa|\eta\> \; \; = \; \; 0,
\en
reveals that the fiducial vectors in (15) form a continuous family (over the parameters $\alpha$ and $\beta$) of \it minimum uncertainty states\rm, satisfying equality in the general affine uncertainty relation
\bn
\<\Delta\sigma\>^2 \; \<\Delta\kappa\>^2 &\geq& \<\sigma\>^2/4.
\en
It is straightforward to show that the associated coherent state overlap function is given by
\bn
\<F', B'|F,B\> = \left [ \frac {e^{-(B'+B)/2}} {(e^{-B'} + e^{-B})/2 + i(F'-F)/2\beta} \right ] ^{2\alpha + 1}.
\en\\

For future comparison, it is helpful to restate our results in terms of the natural variable $G \equiv e^{B}$.  The overlap function then takes the form
\bn
\<F', G'|F,G\> = \left [ \frac {{G'}^{-1/2} G^{-1/2}} {({G'}^{-1} + G^{-1})/2 + i(F'-F)/2\beta} \right ] ^{2\alpha + 1},
\en
while the resolution of unity may be written as an integral over a suitable flat measure:
\bn
N^{-1}\int_{-\infty}^{\infty} dF \int_{0}^{\infty} \; dG \;|F,G\>\<F,G| \;  = \; \one.
\en
The polarization property now appears as
\bn
\left[ iG^{-1}\frac{\partial}{\partial F} \; - \; \beta^{-1}G \frac{\partial}{\partial G} \; - \; \gamma_1 \right] \<F, G|\psi\> \; = \; 0,
\en
where we have introduced
\bn
\gamma_1 \equiv \beta^{-1}(\alpha + \half).
\en

\section{Path Integral for 1-D Coherent States}\

Before building a path integral for the coherent state overlap, we introduce two relevant geometrical objects - the {\it symplectic potential} and the {\it ray metric}.  Suppose that a set of normalized vectors $\{|l\>\}$ defines a continuous curve in a Hilbert space, the label $l$ being a continuous real parameter along the curve.  Then the overlap of two nearby vectors $|l-dl/2\>$ and $|l+dl/2\>$ may be expanded as   
\bn
\<l+dl/2 \; | \; l-dl/2\> \; = \exp(i \; d\theta \; - \; d\Sigma^2/2 \; + \; \cdots)
\en
where $d\theta$ and $d\Sigma^2$ are called, respectively, the symplectic potential and the ray metric.\

The expression (28) may be used to expand the overlap of two of the coherent states from (12),
\bn
&& \hskip-13.25cm \nonumber \<F+dF/2 \; , \; B+dB/2 \; | \; F-dF/2 \; , \; B-dB/2 \>\\
= \exp \; \left\{-i  \;  (\alpha + \half) \; \beta^{-1}e^B \; dF \; - \; (\alpha + \half) \; [\beta^{-2} \; e^{2B} \; (dF)^2 \; +\; (dB)^2] \right\}.
\en
The symplectic potential and the ray metric may now be read off,
\bn
\nonumber d\theta &=& -\gamma_1 \; e^B \; dF\\
&=& -\gamma_1 \; G \; dF\\
\nonumber d\Sigma^2 \; &=& \; \gamma_1 \; [\beta^{-1} e^{2B} \; (dF)^2 \; + \beta \; (dB)^2]\\
&=& \gamma_1 \; [\beta^{-1} G^2 \; (dF)^2 \; + \; \beta G^{-2} \; (dG)^2].
\en

We now introduce a general time-independent Hamiltonian operator ${\cal H}$ with ``upper symbol", $H(F, G)$, defined by
\bn
H(F, G) = \<F, G|{\cal H}|F, G\>,
\en
and ``lower symbol", $h(F, G)$, defined implicitly by
\bn
{\cal H} = N_1^{-1} \int_{-\infty}^{\infty} dF \int_0^{\infty} \; dG \; h(F, G) \; |F, G\>\<F, G|,
\en
and outline the standard construction of the coherent state path integral for the propagator
\bn
J_T(F'', G''; F', G') \equiv \<F'', G''|e^{-i{\cal H}T}|F', G'\>.
\en
The procedure starts with the insertion of $M$ resolutions of unity,
\bn
&& \hskip-11cm \nonumber J_T(F'', G''; F', G')\\ 
= N^{-M} \int \prod_{j=1}^{M} dF_j \; dG_j \prod_{k=0}^M \<F_{k+1}, G_{k+1}|e^{-i\frac{t}{M+1}{\cal H}}|F_{k}, G_{k}\>,
\en
where
\bn
|F_0, G_0\> &\equiv& |F', G'\>,\\
|F_{M+1}, G_{M+1}\> &\equiv& |F'', G''\>.
\en
The $M\rightarrow\infty$ limit is then taken.  It is customary (though not rigorously justified) at this point to interchange the order of the limit and the integrations, and to write the integrand in the form it would take for continuous and differentiable paths.  One is then led, with the aid of the symplectic potential (30), to the strictly formal expression
\bn
J_T(F'', G''; F', G') = {\cal M} \int \exp \left\{i\int_0^T dt \; [-\gamma_1 G {\dot F} - H(F, G)] \right\} \; {\cal D}F \; {\cal D}G,
\en
where ${\cal M}$ represents a suitable normalization.\

An alternative path integral representation may be given using the technique of {\it continuous time regularization} \cite{dkp}.  This procedure involves the insertion of an appropriate Wiener measure into the integral, and leads to
\bn
\nonumber J_T(F'', G''; F', G') = \lim_{\nu \rightarrow \infty} {\cal M_{\nu}} \int \; \exp \left\{i\int_0^T dt \; [-\gamma_1 G {\dot F} - h(F, G)] \right\}\\
\times \; \; \exp \left\{ -(\gamma_1/2\nu)\int_0^T dt \; [\beta^{-1}G^2{\dot F}^2 \; + \; \beta G^{-2}{\dot G}^2]\right\} \; {\cal D}F \; {\cal D}G.
\en
Note that the lower symbol for the Hamiltonian is involved in this formulation.

\section{A Generalized Affine Algebra}\    

We now construct an $n$-dimensional generalization of (1) via the introduction of the set of $n ^{2}$, $GL(n,R)$ generators $\kappa _a^b$ along with their $\half n(n+1)$ symmetric affine conjugates $\sigma _{jk}$ ($=\sigma_{kj}$), satisfying the commutation relations
\bn
\left [\kappa _a^b, \kappa _j^k \right ] &=& i \; \half \left (\delta _a^k \kappa _j^b - \delta _j^b \kappa _a^k \right ),\\
\left [\sigma_{jk},\kappa _a^b\right ] &=& i  \; \half \left (\delta _j^b \sigma _{ak} + \delta _k^b \sigma _{aj}\right ),\\
\left [\sigma_{ab}, \sigma_{jk}\right ] &=& 0,
\en
where all the indices take on values from $1$ to $n$.  The operators $\kappa _a^b$ and $\sigma _{ab}$ may be contracted with sets of constant coefficients $B _b^a$ and $F^{ab}$ ($= F^{ba})$, respectively, and exponentiated to generate the following transformations:
\bn
e^{iB _b^a \kappa _a^b} \; \kappa _j^k  \; e^{-iB _b^a \kappa _a^b}&=& \left (S^{-1}\right )_j^p \kappa_p^q \; S_q^k,\\
e^{iB _b^a \kappa _a^b} \; \sigma _{jk} \; e^{-iB _b^a \kappa _a^b} &=& \left(S^{-1}\right )_j^p \sigma_{pq} \left (S^{-1}\right )_k^q,\\
e^{iF^{ab} \sigma_{ab}} \; \kappa_j^k \; e^{-iF^{ab} \sigma_{ab}} &=& \kappa _j^k \; + \; F^{kp} \sigma_{jp},\\
e^{iF^{ab} \sigma_{ab}} \; \sigma_{jk}\; e^{-iF^{ab} \sigma_{ab}} &=& \sigma_{jk}.
\en
Here the matrix $S$ is defined by 
\bn
S = e^{- \frac {1}{2} B },
\en
and clearly we have
\bn
\det S > 0
\en
for all values of $B$.  Observe that (43) and (44) have the flavor of \it coordinate transformations \rm of tensors of the appropriate valences.

\section{Unitary Representation for the Generalized Affine Algebra}\

We now construct a unitary representation for the group generated by the generalized affine algebra described in section 4. As a representation space we choose the space of square integrable functions of a symmetric, positive-definite matrix variable \rm $\underline{k} \equiv \{k_{ab}\}$, endowed with the inner product definition

\bn
\<\phi \;|\; \psi\> \; = \int_+ \prod _{a\leq b}dk_{ab} \; \phi(\underline{k})^*\; \psi (\underline{k}), 
\en
where the ``+" as a limit to the integral indicates that the domain of integration extends only over those regions in which $\{k_{ab}\}$ is positive definite. The algebra may be represented as follows:
\bn
\sigma _{ab} &=& k_{(ab)},\\
\nonumber \kappa _a^b &=& -i \; \half [\partial ^{(bp)}k_{(pa)} + k_{(pa)}\partial ^{(bp)} ]\\
&=& -i[k_{(ap)}\partial^{(bp)} +  \; \quarter (n+1) \delta _a^b ],
\en
where
\bn
k_{(ab)} &\equiv& \half (k_{ab} + k_{ba}),\\
\partial^{(ab)} &\equiv& \half \left(\partial/\partial k_{ab} + \partial/\partial k_{ba}\right).
\en
It follows from (51) that
\bn
\nonumber e^{-iB _q^p \kappa _p^q} \; \psi (\underline{k}) &=& (\det S)^{(n+1)/2} \; e^{-B _q^p k_{pr}\partial ^{qr}}\psi (\underline{k})\\
&=& (\det S)^{(n+1)/2} \; \psi (S^T \underline{k} S),
\en
where
\bn
(S^T \underline{k} S)_{ab} \equiv S_a^p k_{pq} S_b^q.
\en
It may be verified that with the choice of measure in (49), the operators representing $\sigma_{jk}$ and $\kappa _a^b$ are self adjoint and that the representation of the relevant group is thus rendered unitary:
\bn
|\!| \; e^{-iB_q^p \kappa _p^q}  \; |\psi \> \;  |\!|^2 &=& |\!| \; |\psi \> \;  |\!|^2;\\
|\!| \; e^{-iF^{jk} \sigma_{jk}}  \; |\psi \> \;  |\!|^2 &=& |\!| \; |\psi \> \;  |\!|^2.
\en

\section{Generalized Affine Coherent States}\

Following the procedure in section (2), we now define a family of unitary operators $U(F, S)$ via  
\bn
U(F, S) =  e^{iF ^{jk} \sigma _{jk}} \; e^{-iB _b^a \kappa _a^b}. 
\en
The composition rule for the operators in (\theequation) is
\bn
U(F', S')\; U(F, S)= U(F' + S'^T FS' \; , \; S'S).
\en
where
\bn
(S'^T F S')^{jk} &\equiv& {S'}^j_p F^{pq} S'^k_q,\\
(S'S)^b_a &\equiv& {S'}^p_a S^b_p.
\en
This family of unitary operators may be used to construct a set of coherent states:
\bn
|F, S \> \equiv U(F, S) \; |\eta\>.
\en
Here $|\eta\>$ is an as yet unspecified normalized fiducial vector in the $n$-dimensional representation space.

A resolution of unity in terms of the coherent states in (\theequation) may be established in the usual way, namely, by integrating coherent state projection operators weighted with the appropriate group invariant measure:
\bn
N^{-1}\int_{-\infty}^{\infty} \; \prod_{j \leq k} \; dF^{jk} \; \int_{\det S > 0} \; \prod_{a, b} dS_a^b \; (\det S)^{-(2n + 1)} \; |F, S\>\<F, S| = \one,
\en
the validity of which hinges on the fiducial vector admissibility criterion
\bn
\nonumber N &\equiv& (2\pi)^{n(n+1)/2} \int_{\det S > 0} \prod _{a, b}dS_a^b \; (\det S)^{-n} \; |\eta(S S^T)|^2\\
&=& 2^{-1} (2\pi)^{n(n+1)/2} \int \prod _{a, b}dS_a^b \; |\det S|^{-n} \; |\eta(S S^T)|^2  \; < \; \infty,
\en
where
\bn
(S S^T)_{ab} \equiv \sum_p S_a^p S_b^p.
\en
We now choose as a fiducial vector a natural generalization of the one-dimensional vector in (15), that is,
\bn
\nonumber \eta (\underline{k}) &=& C_n(\alpha , \beta) \; \det (\underline{k}^\alpha e^{-\beta \underline{k}})\\
&=& C_n(\alpha , \beta) \; (\det \underline{k})^\alpha \; e^{-\beta \; {\rm tr} \; \underline{k}},
\en
where $\alpha$ and $\beta$ are positive real coefficients, and the constant $C_n(\alpha , \beta)$ is chosen to be
\bn
C_n(\alpha , \beta) = \frac{(2\beta)^{\alpha n \; + \; n(n+1)/4}}{\sqrt{K_n(2\alpha)}}.
\en
Here $K_n(2\alpha)$ is defined by
\bn
K_n(2\alpha) \; \equiv \; \int_+\; \prod _{a\leq b}dk_{ab} \; (\det \underline{k})^{2\alpha} \; e^{-{\rm tr} \; \underline{k}} .
\en
The integral in (\theequation) may be reduced to a Gaussian integral via a change of variables involving the replacement 
\bn
k_{ab} = \sum_p Q_a^p Q_b^p.
\en
The result is:
\bn
K_n(2\alpha) = 2^{n-1} \; \Omega_n^{-1} \int \prod_{a,b}{dQ_a^b} \; |\det Q|^{4\alpha + 1} \; \exp[-\sum_{a,b}{(Q_a^b)^2}].
\en
Here $\Omega_n$ is the group volume of $SO(n)$,  which can be expressed as a product of the surface volumes of $j$-spheres \cite{gilmore}:
\bn
\Omega_n \; = \; \prod_{j=1}^n  \; \frac{2\pi^{j/2}}{\Gamma(j/2)}.
\en
The change of variables necessary to obtain (70) is used repeatedly throughout this paper - we refer the reader to the appendix for the details.  Clearly the existence (convergence) of the integral expression for $K_n$ is manifest in the form (70).   The choice (67) ensures that all the coherent states are normalized,
\bn
\<F, S | F,S \> = \<\eta | \eta\> = 1.
\en
The admissibility of the fiducial vector (66) may be verified by demonstrating the existence of the integral in (64).  Again we refer the reader to the appendix for the change of variables necessary to perform this type of integral; the result is
\bn
N &=& 2^{-n} \; (4 \pi \beta)^{n(n+1)/2} \; \Omega_n \; \frac{K_n\left(2 \alpha - (n+1)/2\right)}{K_n(2\alpha)} \; < \; \infty.
\en
The overlap of two coherent states based on the fiducial vector (66) may be expressed as
\bn
&& \hskip-12.5cm \nonumber \<F', S' \ | F, S\>\\
 = [C_n(\alpha, \beta)]^2 \; [\det(S'S)]^{(n+1)/2 \; + \; 2\alpha} \int_+ \prod _{a\leq b}dk_{ab} \; (\det \underline{k})^{2\alpha} \; e^{-{\rm tr} (X\underline{k})},
\en
where the complex symmetric matrix $X$ is defined by
\bn
X \equiv \beta(S'^T S' + S^TS) + i(F' - F),
\en
with
\bn
(S^T S)^{ab} &=& \sum_p S_p^a S_p^b,\\
(S'^T S')^{ab} &=& \sum_p {S'}_p^a {S'}_p^b.
\en
The $X$-dependence may be extracted from the integral in (74) to leave
\bn
\nonumber && \hskip-11cm \<F', S' \ | F, S\>=[C_n(\alpha, \beta)]^{2} \left[ \frac{\det(S^{'}S)}{\det X}\right ] ^{(n+1)/2 \; + \; 2\alpha} K_n(\alpha)\\
=\left \{ \frac {\det(S'S)}{\det[(S'^T S' + S^TS)/2 \; + \; i(F' - F)/2\beta]} \right \}^{2\alpha \; + \; (n+1)/2}.
\en
We appeal to analytic continuation to give the final result in (\theequation) a well-defined meaning.

It will be noticed that the overlap function (\theequation) only depends on $S$ through the symmetric combination $S^T S$. It is therefore invariant under a transformation
\bn
S\rightarrow MS,
\en
where $M$ is any $SO(n)$ matrix.  It is appropriate, then, to view the $SO(n)$ degrees of freedom as ``gauge" degrees of freedom and factor them out of the representation.  To this end, we define a new symmetric matrix variable $G$ via the relations
\bn
G^{ab} &\equiv& (S^T S)^{ab} \; \equiv \; \sum_p S_p^a S_p^b,\\
G_{ap} G^{pb} &\equiv& \delta_a^b,\\
G &\equiv& \{G_{ab}\},
\en
and label our coherent states with the parameters $F$ and $G$.  The overlap function (78) then reads
\bn
\<F', G' | F, G\>&=&\left \{ \frac {(\det G')^{-1/2} (\det G)^{-1/2}}{\det \left [({G'}^{-1} + G^{-1})/2 \; + \; i(F' - F)/2\beta \right ] } \right \}^{2\alpha \; + \; (n+1)/2}.
\en
The $SO(n)$ variables may be integrated out of the resolution of unity (again, see the appendix for a discussion of the required change of variables), the result being
\bn
2^{-n} N^{-1} \Omega_n \int \prod_{j\leq k} dF^{jk} \int_+ \prod_{a\leq b} dG_{ab} \; |F, G\>\<F, G| &=& \one.
\en
We note that the removal of the $SO(n)$ degrees of freedom from the representation is only appropriate if the dynamics in question is governed by a classical Hamiltonian which is a function of $F$ and $G$; this is the point of view we shall take in the remainder of this article.  It is, however, easy in principle to envisage Hamiltonians where spinor-like variables couple directly to $S$, in which case it would of course be necessary to retain the label $S$.\

The polarization property analogous to (26) for the one-dimensional case may be written as
\bn
\left\{ iG^{ap}\frac{\partial}{\partial F^{pb}} \; + \; \beta^{-1} G^{ap} \frac{\partial}{\partial G^{pb}} \; - \; \gamma \delta_b^a \right\} \< F, G|\psi \> \; = \; 0,
\en
where we have written 
\bn
\gamma \equiv \beta^{-1} [(n+1)/4 + \alpha].
\en

\section{Path Integral for the Propagator}\

The procedure in section 3 may be followed to build a formal path integral expression for the $n$-dimensional propagator associated with a general time-independent Hamiltonian ${\cal H}$ with upper symbol
\bn
H(F, G) = \<F, G|{\cal H}|F, G\>.
\en
We first construct the relevant symplectic potential and ray metric.  The identity
\bn
\nonumber \det(1+dA) &=& e^{{\rm tr} \ln (1 + dA)}\\
&=& e^{{\rm tr}(dA \; - \; dA^2/2 \; + \; \cdots)}
\en
where $dA$ is any infinitesimal matrix, may be used to expand the overlap of two neighboring coherent states as follows:
\bn
\<F+dF/2 \; , \; S+dS/2 \; | \; F-dF/2 \; , \; S-dS/2 \> \; = \; e^{i \; d\theta \; - \; d\Sigma^2/2},
\en
where the 1-form $d\theta$ is given by
\bn
d\theta = -\gamma \; {\rm tr}(G \; dF),
\en
and the ray metric $d\Sigma^2$ by
\bn
d\Sigma^2 \; = \; \gamma \; \{\beta \; {\rm tr}[(G^{-1}dG)^2] \; + \; \beta^{-1}{\rm tr}[(G \; dF)^2]\}.
\en
The propagator may then be written as
\bn
&& \hskip-12.5cm \nonumber J_T(F'', G''; F', G') \equiv \<F'', G''|e^{-i{\cal H}T}|F', G'\>\\
={\cal M} \int \prod_{j\leq k} {\cal D}F^{jk} \prod_{a\leq b} {\cal D}G_{ab} \exp {\left\{i\int_0^T [-\gamma \; {\rm tr}(G {\dot F}) - H(F, G)] dt\right\}},
\en
a strictly formal expression to which the remarks immediately preceding (38) again apply.\

An alternative representation for the propagator which uses a continuous-time regularization and the lower symbol is given by
\bn
\nonumber && \hskip -13cm J_T(F'', G''; F', G') \\ 
\nonumber = \lim_{\nu \rightarrow \infty} {\cal M_{\nu}} \int \prod_{j\leq k} {\cal D}F^{jk} \prod_{a\leq b} {\cal D}G_{ab} \; \exp \left\{i\int_0^T[-\gamma \; {\rm tr}(G {\dot F}) - h(F, G)]dt \right\}\\
\times \; \; \exp \left\{ -(\gamma /2\nu) \int_0^T dt \; \{\beta \; {\rm tr}[(G^{-1}{\dot G})^2] \; + \; \beta^{-1}{\rm tr}[(G \; {\dot F})^2]\}\right\}.
\en

\section{Conclusion}\

In this article we have constructed a framework for the quantization of a positive-definite matrix degree of freedom $\{\sigma_{ab}\}$.  Specifically, we have demonstrated that complementing the operators $\sigma_{ab}$ with {\it affine} conjugates (40-42) leads to a representation in which the spectrum of the matrix operator $\{\sigma_{ab}\}$ is strictly  positive definite.  Such an approach appears far more satisfactory than the standard use of canonical commutation relations, where the positivity of $\{\sigma_{ab}\}$ can only be insured by rather artificial means, if at all.  As demonstrated, the generalized affine algebra leads to a set of group-defined coherent states that have been used to construct two versions of path integral expressions for the propagator.  Finally, we suggest that the affine construction we have outlined is well suited to the quantization of the spacial part of the metric tensor of general relativity, a program already initiated in \cite{klauder2}.

\pagebreak

\section*{Appendix - \\ Jacobian associated with change of variables from $S$ to $G$}\

We have repeatedly found it useful to factor out the $SO(n)$ degrees of freedom from a real $n$-dimensional non-degenerate matrix with positive determinant $S$ and display the remaining degrees of freedom as elements of the matrix $G \equiv S^T S$.  We now derive the form of the Jacobian associated with this change of variables.\

We first calculate the Jacobian associated with the polar decomposition of $S$.  Such a decomposition may be defined by 
\bn
S=MT,
\en
where $M \in SO(n)$ and $T$ is a positive $n$-dimensional upper triangular matrix.  We first consider the case $n=2$, writing (\theequation) explicitly as
\bn
\left(\matrix{S_1^1 & S_1^2 \cr
        S_2^1 & S_2^2 \cr}\right) = \left(\matrix{\cos \theta & -\sin \theta\cr
                                                    \sin \theta & \cos \theta\cr}\right) \left(\matrix{T_1^1 & T_1^2 \cr
                                                                                                        0 & T_2^2 \cr}                                                                                                                     \right).
\en
It is straightforward to show that in this case,
\bn
dS_1^1 \wedge dS_1^2 \wedge dS_2^1 \wedge dS_2^2 \; = \; T_1^1 \; d\theta \wedge dT_1^1 \wedge dT_1^2 \wedge dT_2^2.
\en
We now generalize (\theequation) for $n>2$.  It is expeditious to express the $SO(n)$ matrix $M$ as a product of $\half n(n-1)$, $SO(n)$ matrices $R_{ij}={R_{ij}(\theta_{ij})}$, $n\geq i>j$, each of which represents a rotation about the $ij$ axis through an angle $\theta_{ij}$:
\bn
\nonumber &&\hskip-9cm M = (R_{21} \; R_{31} \; \cdots \; R_{n1}) \; (R_{32} \; R_{42} \cdots \; R_{n2})\\
\times \cdots \times (R_{(n-1)(n-2)} \; R_{n(n-2)}) \; (R_{n(n-1)}).
\en
It will be noticed that $R_{n(n-1)}$, whose explicit form is  
\bn
R_{n(n-1)} \; = \; \left( \begin{array}{cccccc}
1 & 0 & 0 & \cdots & \cdots & 0 \\
0 & 1 & 0 & \cdots & \cdots & 0 \\
0 & 0 & 1 & \cdots & \cdots & 0 \\
\vdots & \vdots & \vdots  & \ddots & & \\
0 & 0 & 0 &       & \cos \theta_{n(n-1)} & -\sin \theta_{n(n-1)} \\
0 & 0 & 0 &       & \sin \theta_{n(n-1)} & \cos \theta_{n(n-1)} \\
\end{array} \right),
\en
only affects the bottom right $2 \times 2$ block of $T$.  It is therefore responsible for the introduction of a factor $T_{(n-1)}^{(n-1)}$ into the Jacobian.  Similarly, each $R_{ij}$ only acts on the bottom $j \times j$ block of $T$, introducing a factor $T_j^j$.  Building up the entire Jacobian in this way, we find that
\bn
dS_1^1 \wedge \cdots \wedge dS_n^n \; = \; (T_1^1)^{\; n-1} \; (T_2^2)^{\; n-2} \; \cdots \; (T_{n-1}^{n-1}) \; d\Omega \wedge dT_1^1 \wedge \cdots \wedge dT_n^n
\en
where $d\Omega$ is the invariant measure on $SO(n)$. 

Having separated the matrix $S$ into its ``radial" part $T$ and ``angular" part $M$, we are now in a position to make a change of variable from $T$ to $G$ (recall that both of these matrices possess $\half n(n+1)$ degrees of freedom):
\bn
G = S^T S = T^T T.
\en
Inspection of the elements of $G$ quickly reveals the form of the Jacobian associated with this change of variable:
\bn
\nonumber && \hskip-11.25cm dG^{11} \wedge \cdots \wedge dG^{nn}\\
= \; 2^n \; (T_1^1)^{\; n} \; (T_2^2)^{\; n-1} \; \cdots \; (T_{n-1}^{n-1})^{\; 2} \; (T_n^n) \; dT_1^1 \wedge \cdots \wedge dT_n^n.
\en

Combining (99) and (\theequation), and noting that $\det S = \det T$, we obtain our central result:
\bn
d\Omega \wedge dG^{11} \wedge \cdots \wedge dG^{nn} &=& 2^n \; (\det S) \; dS_1^1 \wedge \cdots \wedge dS_n^n
\en
We now integrate a general function $f(G)$ against the measure in (\theequation):
\bn
\int _+ \prod_{a \leq b} dG^{ab} \; f(G) = \; 2^n \; \Omega _n ^{-1} \; \int_{\det S > 0} \prod_{a,b} \; dS_a^b \; (\det S) \;f(S^T S) ,
\en
where $\Omega_n$ is the group volume of $SO(n)$ given in (71)\cite{gilmore}.

\pagebreak

\end{document}